\documentstyle[11pt,aaspp4]{article}
\begin{document}
\slugcomment{To be published in {\it Astrophys. J. Letters}, 
10 Sept 1997 issue.}

%\tighten

% These are Knox's private definitions

\def\LL{\mbox{$\:\lambda\lambda $ }}
\def\HA{{H$\alpha$ }}
\def\HB{{H$\beta$ }}

%\def\FLUXARCSEC '\:ergs\:cm sup -2 s sup -1 arcsec sup -2'
%\def\COUNTS '\:s sup -1 '
%\def\CPM '\:s sup -1 arcmin sup -2'

%Needs trailing \ except at end of sentence
\def\BT{B_T}
\def\MBT{M_{B_T}}
\def\Mstar{M^*}
\def\Lstar{L^*}
\def\Mpc{\:\rm{Mpc}}
\def\kpc{\:\rm{kpc}}
\def\pc{\:\rm{pc}}
\def\kms{\:\rm{\,km\,s^{-1}}}
\def\rfid{r_{\rm{fid}}}

\def\LUM{\:{\rm ergs\:s^{-1}}}
\def\FLUX{\:{\rm ergs\:cm^{-2}\:s^{-1}}}
\def\FLUXARCSEC{\:{\rm ergs\:cm^{-2}\:s^{-1}\:arcsec^{-2}}}
\def\FLUXARCMIN{\:{\rm ergs\:cm^{-2}\:s^{-1}\:arcmin^{-2}}}
\def\FLUXSR{\:{\rm ergs\:cm^{-2}\:s^{-1}\:sr^{-1}}}
\def\COUNTS{\:{\rm counts\:s^{-1}}}
\def\CPM{\:{\rm counts\:s^{-1}\:arcmin^{-2}}}
\def\VEL{\:{\rm km\:s^{-1}}}
\def\ETAL{{\it et\:al.}}
\def\OIGS{\:{\rm ergs\:cm^{-2}\:s^{-1}\:\AA^{-1}}}
\def\LA{Lyman\thinspace$\alpha$}

%A few moreprivate definitions
\def\eg{{\it e.g.}}
\def\ie{{\it i.e.}}
\def\etal{{\it et\thinspace al.}\ }
\def\ergflux{ergs\thinspace cm$^{-2}$\thinspace s$^{-1}$\ }
\def\ha{H$\alpha$}
\def\hb{H$\beta$}

%Use Roman numbers to designate ionization state and first two digits
%of wavelength in Ang. where necessary
\def\oiii{[\ion{O}{3}]}
\def\fei{\ion{Fe}{1}}
\def\feii{\ion{Fe}{2}}
\def\feiii{\ion{Fe}{3}}
\def\siII{\ion{Si}{2}}
\def\siIII{\ion{Si}{3}}
\def\siIV{\ion{Si}{4}}
\def\feplus{Fe$^+$}

\def\HEiiL{\ion{He}{2} $\lambda1640$}
\def\NvL{\ion{N}{5} $\lambda\lambda1239,1243$}
\def\CivL{\ion{C}{4} $\lambda\lambda1548,1551$}
\def\OviL{\ion{O}{6} $\lambda\lambda1032,1038$}
\def\SIivL{\ion{Si}{4} $\lambda\lambda1394,1403$}

%private definitions end here

% Additional private definitions that appear to work only inside document

\newcommand{\MSOL}{\mbox{$\:M_{\sun}$}}  

\newcommand{\EXPN}[2]{\mbox{$#1\times 10^{#2}$}}
\newcommand{\EXPU}[3]{\mbox{\rm $#1 \times 10^{#2} \rm\:#3$}}  % exponent with units

% End of defining things

% End of defining things

\title{A Second UV ``Light Bulb'' behind the SN~1006 Remnant}

\author{P. Frank Winkler\altaffilmark{1}}
\affil{Department of Physics, Middlebury College, Middlebury, VT  05753}
\authoraddr{email: winkler@middlebury.edu}

\and

\author{Knox S. Long}
\affil{Space Telescope Science Institute, 3700 San Martin Drive,
Baltimore, MD 21218}
\authoraddr{email: long@stsci.edu}

\altaffiltext{1}{Visiting Astronomer, Cerro Tololo Inter-American Observatory.
CTIO is operated by AURA, Inc.\ under contract to the National Science
Foundation.}

\begin{abstract}
A point X-ray source located 9 arcmin northeast of the center of SN~1006
has been spectroscopically identified as a background QSO, with a redshift
of 0.335.  The object is moderately bright, with magnitude $V=18.3$.
If its ultraviolet spectrum is typical of low-$z$\ quasars, 
this object will be a
second (after the Schweizer-Middleditch star) source to use for 
absorption spectroscopy of material within SN~1006.  Absorption spectra
provide a unique probe for unshocked ejecta within
this supernova remnant, and can possibly solve the long-standing problem of ``missing'' iron in the remnants of Type Ia supernovae.

\end{abstract}


%\twocolumn
\section{Introduction}

As the remnant of the brightest supernova to have been witnessed in
recorded history, SN~1006 has long held interest for astronomers 
working at a variety of wavelengths using varied techniques.  Among
the most powerful of these has been the measurement of absorption
lines in the ultraviolet spectrum of a hot subdwarf star (commonly referred to
as the ``SM star''), discovered and 
classified as sdOB by Schweizer and Middleditch (1980).  
The fortuitous location of this star behind SN~1006, only 2\farcm 5 
from the projected geometric center of the remnant, provides an unusual
opportunity to study the distribution of cold SN ejecta through absorption
spectroscopy. 
Very broad UV resonance absorption features due to \feii\ were 
first detected in IUE spectra of the SM star by Wu \etal (1983; see also
Fesen \etal 1988).  Subsequent spectra from the FOS on {\it HST} 
led to positive identification of additional broad absorption lines of
\siII\ and \siIV, in addition to greatly improved measurements of the 
\feii\ lines (Wu \etal 1993; 1997).

Iron in SN~1006 is of special interest, since it is the suspected remnant
of a Type Ia event (Minkowski 1966; Schaefer 1996), and conventional
models for SN~Ia require the eventual production of several tenths of 
a solar mass of iron---formed from the decay of the $^{56}$Ni 
which powers the SN~Ia light curve 
(Colgate \& McKee 1969; Arnett 1979; Nomoto, Thielemann, \& Yokoi 1984; 
an extensive set of models and references to more recent literature is given
by H\"{o}flich \& Khokhlov 1996).  
However, the \feii\ absorption line measurements indicate only 0.014 \MSOL\
of \feplus, short of the expected total Fe mass by a factor of 20 or more.  
A new analysis of the FOS data using a different method of continuum
fitting by Hamilton \etal (1997) leads to 0.029 \MSOL\ of \feplus, but is still far short
of predictions for the total mass of Fe. In other young remnants of 
probable Type Ia SN, Tycho and Kepler, strong X-ray lines from highly-ionized 
iron result in large part from ejecta that have been excited by a reverse shock,
but the absence of such lines in SN~1006 indicates that iron in the ejecta
cannot yet have been shocked, so must still be relatively cool.  
Hamilton \& Fesen (1988) 
considered various states in which iron could be hidden:  the absence of \fei\ absorption
features excludes the possibility of significant gaseous neutral 
iron, while the 
absence of strong IR emission provides evidence against the formation of
grains (iron-rich or otherwise) in the ejecta.  They suggested that a
significant amount of Fe might be more highly ionized, but far-UV spectra 
of the SM star 
obtained by  Blair, Long, \& Raymond (1996) from {\it HUT} show only
weak \feiii\ absorption and severely limit the amount of Fe$^{++}$.  
Where several tenths of a solar mass of iron might be hiding in SN~1006 
remains a mystery.

Further UV absorption spectroscopy of SN~1006, especially probing different
lines of sight through the remnant shell, would certainly be important,
but one is, of course, limited by the availability of background UV
``light bulbs'' which are appropriately placed and bright enough for 
observation.  We report here
the discovery of a second, moderately bright UV source within SN~1006:  a 
QSO located 9\arcmin\ NE from the projected center of the remnant
shell.  

\section {Observations and Results}

In an observation of SN~1006 from the {\it ROSAT} HRI, we noted 
an unresolved X-ray source, located in the NE region of the remnant shell, with
no obvious optical counterpart (Winkler \& Long 1997; hereafter Paper I).  (In addition,
a second, brighter, X-ray source near the center of the remnant shell, as well as three
fainter sources outside, all coincide with bright, $V \sim$\ 9--10, 
foreground stars.)  The unidentified NE source is also apparent in
the PSPC images of Willingale \etal (1996), especially the image
at $E > 1.5$\ keV, indicating that it is a relatively hard source.
The HRI count rate was measured at $5\ {\rm counts\; ksec}^{-1}$,
equivalent to $\sim 2\times 10^{-13}\;$ \ergflux (0.5--2.5 keV) for a 
hard-spectrum source.
In the \ha\ image from Paper I, the only evident object within the 5\arcsec\ 
error radius for the NE X-ray source was what appeared to be a star estimated at
about magnitude 17.  Located at 
${\rm RA}\: (2000) = 15^{\rm h}\, 03^{\rm m}\, 33\fs 93,\ 
{\rm Dec}\: (2000) = -41\arcdeg\, 52\arcmin\, 23\farcs 7$, this object
is only 3\farcs 7 from the position of the HRI X-ray source, well within the
HRI error circle.

On 1997 March 29, we obtained a spectrum of this
candidate object from the CTIO 1.5m f/7.5 telescope and R-C spectrograph.  
The spectrograph was configured with a 300 line~mm$^{-1}$ grating
blazed at 4000 \AA\ and a Loral $1200 \times 800$\ pixel CCD, to give 
wavelength coverage of 3500--7000 \AA\ at a dispersion of 2.9 \AA~pixel$^{-1}$.
A spectrograph slit width of 3\farcs5 gave a resolution of 9 \AA\@.  
The candidate was observed under photometric but brightly moonlit
conditions for a total of 4000 s, split among 4 frames of 1000 s each.  
All the data reduction has been carried out using conventional IRAF\footnote
{IRAF is distributed by the National Optical Astronomy Observatories,
operated by AURA, Inc., under contract from the National Science Foundation.}
reduction techniques:   flat-fielding using a combination of dome flats
and internal quartz flats, secondary ``illumination correction'' using 
twilight sky flats, and wavelength calibration from an internal He-Ar source.
Flux calibration was achieved by observing several spectrophotometric
standards from Hamuy \etal (1992).  

The extracted spectrum is shown in Figure 1.  The relatively flat
spectrum and pattern of broad and narrow emission lines is typical of
QSOs.  The strongest line is \ion{Mg}{2} $\lambda\,2798$, redshifted
into the blue end of the optical spectrum with velocity width 
$\sim 7000 \kms$\ (FWHM)\@.  Also evident are broad \hb\ and narrow
\oiii\ $\lambda\lambda\, 4959,\, 5007$, at the red end of the spectrum.
A redshift $z = 0.335 \pm 0.001$\ is consistent with all the identified features.
%\notetoeditor{Figure 1 goes about here.  We recommend slightly larger
%than 1-column format:  about 4 in. wide x 3 in. high, but it should be
%OK in single-column format if space is limited.}

We have also obtained broad-band images of the NE region
of SN~1006 from the CTIO 0.9 m telescope, equipped with
the Tektronix No. 5 $2048 \times 2048$\ pixel CCD, 
on 1997 February 10.  This combination gives a field of 13\farcm 7 at a 
scale of 0\farcs 40 pixel$^{-1}$\@. Exposures of 1200, 240, and 240 s, respectively,
were obtained in the $U,\ B,$\ and $V$\ bands under photometric, moonless
conditions.  These were reduced using standard IRAF procedures for 
overscan subtraction and flat-fielding based on well-exposed twilight sky flats.  
Photometric calibration was based on exposures of several Landolt (1992) and 
Graham (1982) fields.  Figure 2 shows a section of the $V$\ band image 
with the error circle for the HRI X-ray source indicated.  The QSO position 
given above was measured from this image, using 18 surrounding 
stars from the {\it HST} Guide Star Catalog to define the reference frame.
%\notetoeditor{Figure 2 goes about here, as a half-tone in the text.  
%We recommend a format about 4--4.5 in square.}

Photometry of the object which was subsequently identified as a background
QSO indicates unusual, very blue, colors:  
$V = 18.32\pm 0.05,\ B-V= 0.17\pm 0.04,\ U-B=-0.78\pm 0.04$\@.
This is significantly fainter than our earlier estimate (paper I) of about 
magnitude 17, based on a narrow-band \ha\ image.  
The difference is most likely due to the fact that the QSO redshift
of 0.335 shifts the wavelength of \hb\ emission into the bandpass of our
\ha\ filter.

\section{Discussion}

The discovery of another quasar at low redshift would be entirely unnoteworthy
were it not for its location.  But through its felicitous placement behind 
the SN~1006 remnant, this object becomes potentially quite valuable as
an ultraviolet continuum source against which to measure absorption lines.
As we noted in Section 1, the one line of sight through
SN~1006 which has so far been probed, \ie, that to the SM star, 
has yielded rich information while deepening the mystery of 
where the iron in this supposedly Type Ia remnant may be hiding.  
The broad absorption lines of \feii\ and \feiii\, measured with the
{\it HST} FOS and {\it HUT}, respectively, demonstrate convincingly that a
significant amount of fast-moving, cold iron is present within the remnant
shell, but the inventory of \feplus\ and Fe$^{++}$\ obtained from 
spherically symmetric models
based on this single line of sight falls short of the $\gtrsim 0.3$\ \MSOL\
of iron predicted from models for Type Ia supernovae by a factor of at least
10, and perhaps 20 or more.  

The absorption line measurements for the SM star lead to another problem
as well:  how to keep ejecta expanding at the velocity indicated by the
widths of the absorption lines within the confines of the remnant shell,
without placing SN~1006 too far away.  As we discussed more fully in 
Paper I, the velocity width for the cold iron, together with the known age
of the remnant, gives a minimum extent for ejecta along the line of sight which is barely
smaller than the transverse dimensions of the outer remnant shell at a distance
of 1.8 kpc---the value inferred from 
proper motions of the optical filaments measured by Long, Blair, \& van den Bergh
(1988), together with the velocity of the shock as measured spectroscopically 
in these same filaments by Laming \etal (1996).  Hamilton \etal (1997) reanalyzed
the absorption data for the SM star and concluded that the near-side, blueshifted 
velocity is smaller than that originally measured by Wu \etal 1993.  With this
result and a geometry which is elongated along the line of sight, they were able
to achieve a self-consistent model for SN~1006.  

Both these problems---the ``missing iron'' and the geometry-distance 
puzzle---may be elucidated by probing the distribution of ejecta along a 
second line of sight through SN~1006.  The 
QSO is well placed for this purpose, located 9\arcmin\ away 
from the remnant center,
$\sim 58$\% of the shell radius in this direction, compared with 2\farcm 5 for the 
SM star (see Figure 3).  Absorption would, of course, be observed only from material
which lies outside the the projected radius to the line of sight, so an ejecta ion species 
concentrated entirely within $\lesssim 58$\% of the shell radius would produce
no absorption.  But for a species which is 
smoothly distributed around a spherical shell of radius $\gtrsim 60$\% of the
shell radius, the line of sight to the QSO
may have a larger optical depth and narrower velocity width than that to the SM star. 
Absorption spectra 
and comparision of line profiles to the two sources 
for all species which have been observed in the 
SM star---\feii, \feiii, \siII, \siIII, and \siIV---would 
be important for constraining the geometry
of the ejecta distribution and the radius of the reverse shock in SN~1006.
\notetoeditor{Figure 3 goes about here, as a half-tone in the text.
We recommend single-column format}

Is the QSO bright enough to measure absorption profiles?  We have redshifted the 
composite spectrum from 101 QSOs observed with the {\it HST} FOS 
(Zheng \etal 1997) to $z=0.335$\ and scaled it to match the QSO spectrum from 
Figure 1 in the region of overlap.   The extinction is unknown, but we
have assumed a value $E(B-V)=0.12$, the same as that measured for 
the SM star (Blair \etal 1996).  The continuum is nearly flat at 
$F_\lambda \sim 3 \times 10^{-16}\OIGS$\ over the range 2200--3500 \AA,
about a factor of 50 fainter than that of the SM star at the wavelengths of
the strong \feii\ lines, $\lambda\: 2383$\ and $\lambda\: 2600$.  
Nevertheless, the QSO is sufficiently bright that high quality
UV spectra can be obtained with modest exposures using the new imaging
spectrograph on the {\it Hubble} Space Telescope.

There is an interesting application for the QSO behind SN~1006 for X-ray
astronomy as well.  The brightest part of the X-ray shell is in the NE, 
passing only 6\arcmin\ away from the QSO.  Since the QSO is
a point source of X-rays, it is potentially an excellent fiducial point for measuring
the position and proper motion of the X-ray shell in this area.  This is 
the same region in which Koyama \etal (1995) argued for a synchrotron
origin for the X-ray emission based on the absence of X-ray lines in the
ASCA spectrum.  In Paper I, we showed a detailed correlation between the
X-ray morphology observed with the {\it ROSAT} HRI and that observed
in radio from the VLA by Reynolds \& Gilmore (1986).  Moffett, Goss,
\& Reynolds (1993) have measured the radio expansion rate for the entire
shell at $0.44 \pm 0.13\; {\rm arcsec \: yr}^{-1}$.  This differs somewhat 
from the proper motions for the
\ha\ filaments of $0.30\pm 0.04\;{\rm arcsec \: yr}^{-1}$\ determined by 
Long \etal (1988), but the optical value
was specifically for the NW filaments, while the radio value was for the
overall expansion of the shell.  It would be interesting to determine 
the X-ray proper motion at a specific shell location.  Measurement of 
the expected differences
of a few arcsec over a baseline of several years would be marginal at best
based on the typical aspect uncertainty of $\sim 5\arcsec$\ for the {\it ROSAT}
HRI, but the presence of a fiducial reference so close to the X-ray shock
front would make measurement of the proper motions feasible.

\acknowledgments

We are grateful to Becky Walldroff for her assistance in carrying out the
spectroscopic observations, and to the staff of CTIO for their 
capable and efficient support.  This work has been supported in part 
by NSF grant
AST-9315967 and NASA grant NAG5-1668, with additional support 
from the W.M. Keck Foundation through
the Keck Northeast Astronomy Consortium.   PFW gratefully acknowledges the 
hospitality of CTIO, where he has been in residence during the course
of the work presented here, and in particular valuable discussions with J.
Baldwin, M. Keane, and M. Phillips.

\newpage
\figcaption [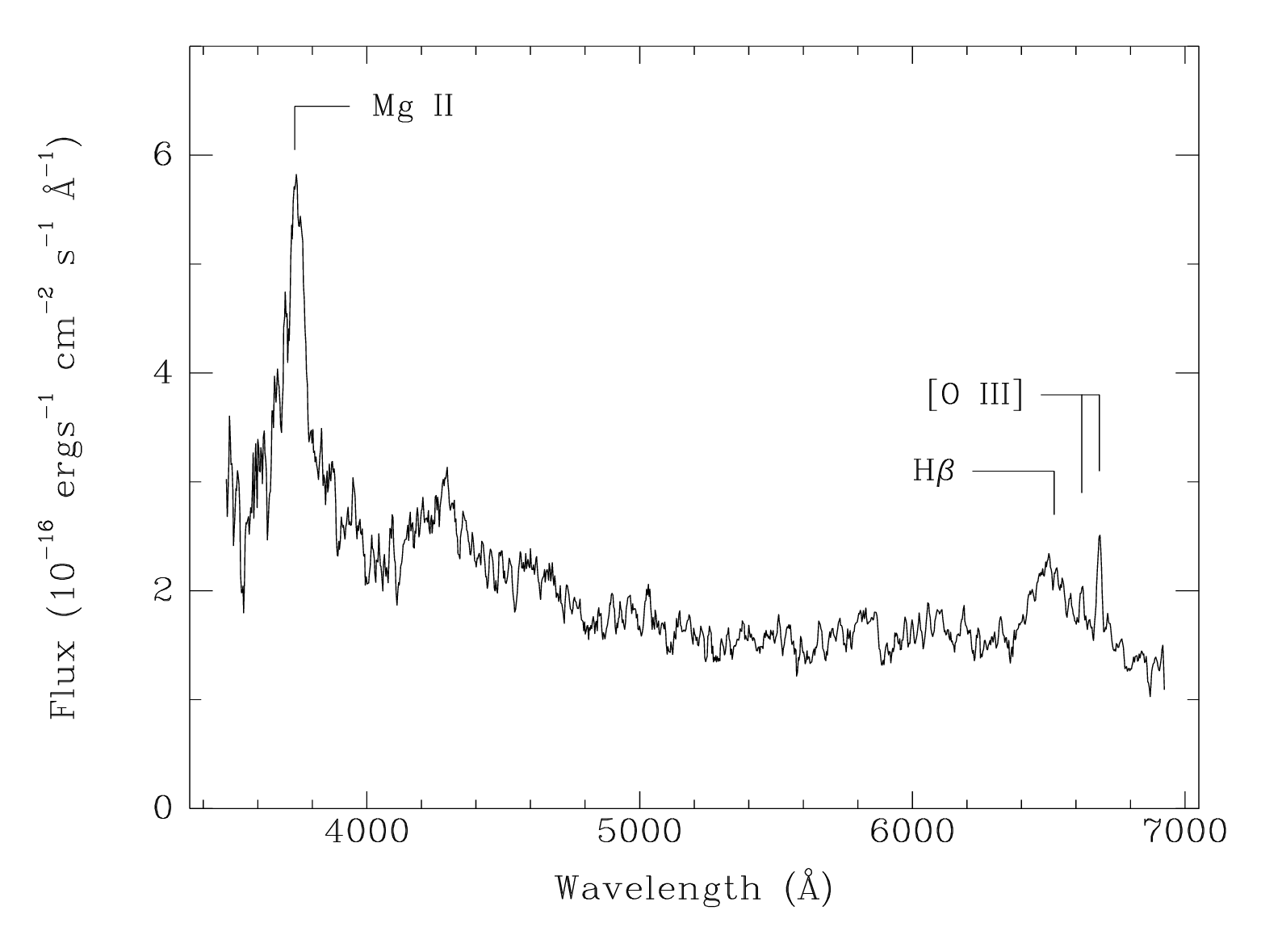]{Optical spectrum of the counterpart to the 
X-ray source shown in Figure 2, now identified as a QSO at redshift 0.335.}
\figcaption [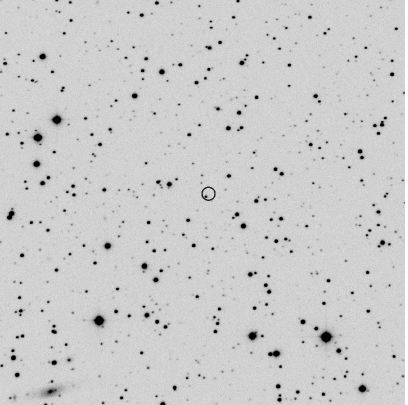]{
A  section of the {\it V} band image of the
NE region of SN~1006,
showing the 10\arcsec\ diameter error circle for the 
X-ray point source noted by Winkler \& Long (1997).  The 
object within the circle is the QSO with the spectrum shown in Figure 1. 
Several background galaxies may also be seen, indicative of the low
extinction in the direction of SN~1006.  The illustrated section measures exactly 
$6\arcmin \times 6\arcmin$; north is up and east is to the left.}

\figcaption [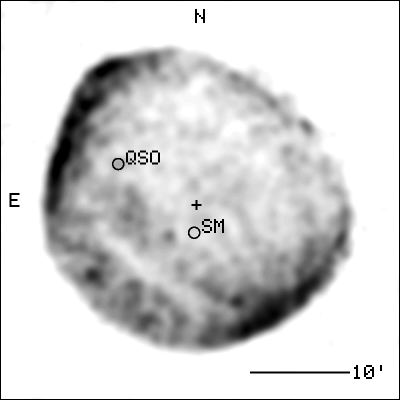]{Smoothed X-ray map 
(from Winkler \& Long 1997), showing the locations
of the newly discovered QSO and the SM star within the SN~1006
shell.  The plus indicates the center
of the remnant.  The frame measures $40\arcmin \times 40\arcmin$.}


\begin{references}
\reference{}
Arnett, W.D. 1979, \apjl, 230, L37
%Theory of SNI:  plausible WD accretion model for SNI based on Ni decay

\reference{}
Blair, W.P., Long, K.S., \& Raymond, J.C. 1996, \apj, 468, 871
%HUT obsn of FeIII in SM star

\reference{}
Colgate, S.A., \& McKee, C. 1969, \apj, 157, 623
%First suggestion of Ni-Co-Fe decay for SNI

%\reference{}
%Cardelli, J.A., Clayton, G.C., \& Mathis, J.S. 1989, \apj, 345, 245
%Best UV/optical/IR extinction curves

\reference{}
Fesen, R. A., Wu, C.-C., Leventhal, M., \& Hamilton, A. J. S. 1988, \apj,
327, 164
%Reanalysis of IUE data on SM star

\reference{}
Graham, J.A. 1982, \pasp, 94, 244

\reference{}
Hamilton, A. J. S., \& Fesen, R.A. 1988, \apj, 327, 178
%Reionization of unshocked ejecta in SN1006

\reference{}
Hamilton, A. J. S.,  Fesen, R. A., Wu C.-C., Crenshaw, D. M., \& Sarazin,
C. L 1997, \apj, 481, 838   

\reference{}
Hamuy, M., Walker, A.R., Suntzeff, N.B., Gigoux, P., Heathcote, S.R., 
\& Phillips, M.M. 1992, \pasp, 104, 533

\reference{}
H\"{o}flich, P. \& Khokhlov, A. 1996, \apj, 457, 500

\reference{}
Koyama, K., Petre, R., Gotthelf, E.V., Hwang, U., Matsuura, M., Ozaki, M., \& Holt, S.S. 1995, \nat, 378, 255

\reference{}
Laming, J.M., Raymond, J.C., McLaughlin, B.M., \& Blair, W.P. 1996, \apj, 472, 267
% electron-ion equilibration in SN1006

\reference{}
Landolt, A.U. 1992, \aj, 104, 340

\reference{}
Long, K.S., Blair, W.P., \& van den Bergh, S. 1988, \apj, 333, 749

\reference{}
Minkowski, R. 1966, \aj, 71, 371
%Concludes that SN1006 was probably Type I.

\reference{}
Moffett, D.A., Goss, W.M., \& Reynolds, S.P. 1993, \aj, 106, 1566

\reference{}
Nomoto, K., Thielemann, F.-K., \& Yokoi, K. 1984, \apj, 286, 644
% First carbon-deflagration models for SNI

\reference{}
Reynolds, S.P., \& Gilmore, D.M. 1986, \aj, 92, 1138

\reference{}
Schaefer, B.E. 1996, \apj, 459, 438
%"Historical SN and the Hubble const" -- reviews distance for SN1006

\reference{}
Schweizer, F., \& Middleditch, J. 1980, \apj, 241, 1039
% Discovery paper for SM star

\reference{}
Willingale, R., West, R.G., Pye, J.P., \& Stewart, G.C. 1996, \mnras, 278, 749
% ROSAT PSPC of SN1006

\reference{}
Winkler, P.F., \& Long, K.S. (1997), \apj, in press (Paper I)

\reference{}
Wu C.-C., Crenshaw, D. M., Fesen, R. A., Hamilton, A. J. S. \& Sarazin,
C. L 1993, \apj, 416, 247   %FOS observation of S-M star

\reference{}
Wu C.-C., Crenshaw, D. M., Hamilton, A. J. S., Fesen, R. A., Leventhal, M. 
\& Sarazin, C. L 1997, \apjl, 477, L53   
%FOS observation of S-M star; far UV

\reference{}
Wu, C.-C., Leventhal, M., \& Sarazin, C. L., \& Gull, T. R. 1983,
\apj, 269, L5
%Original IUE discovery that S-M star had Fe II absorption lines

\reference{}
Zheng, W., Kriss, G.A., Telfer, R.C., Grimes, J.P., \& Davidsen, A.F. 1997,
\apj, 475, 469
%Composite HST FOS spectrum of QSOs

\end{references}
\end{document}